# Low Stress Ion Conductance Microscopy of Sub-Cellular Stiffness†

Richard W. Clarke,‡*[a] Pavel Novak,‡[b] Alexander Zhukov,[a] Eleanor J. Tyler,[c] Marife Cano-Jaimez,[d] Anna Drews,[a] Owen Richards,[a] Kirill Volynski,[d] Cleo Bishop[c] and David Klenerman*[a]

Directly examining subcellular mechanics whilst avoiding excessive strain of a live cell requires the precise control of light stress on very small areas, which is fundamentally difficult. Here we use a glass nanopipet out of contact with the plasma membrane to both exert the stress on the cell and also accurately monitor cellular compression. This allows the mapping of cell stiffness at a lateral resolution finer than 100 nm. We calculate the stress a nanopipet exerts on a cell as the sum of the intrinsic pressure between the tip face and the plasma membrane plus its direct pressure on any glycocalyx, both evaluated from the gap size in terms of the ion current decrease. A survey of cell types confirms that an intracellular pressure of approximately 120 Pa begins to detach the plasma membrane from the cytoskeleton and reveals that the first 0.66±0.09 µm of compression of a neuron cell body is much softer than previous methods have been able to detect.

## Introduction

In Ion Conductance Microscopy (ICM), insulating surfaces in conducting solution are detected by their slight occlusion of the ion current through the tip aperture of a nanopipet probe.[1] A picoampere drop in this nanoampere ion current between the capillary and bath electrodes can be detected within a millisecond using a patch-clamp amplifier, allowing piezoelectric positioning of the nanopipet to map a cell's topography[2] and to patch to an exact point of interest.[3] It was long thought that ICM imaging exerts almost no stress on a cell simply because the feedback control keeps the tip from making contact with it. However, during approaches to cells the ion current decreases far more slowly with height than its rapid drop next to a hard surface, indicating the glass tip face repels the cell membrane before contact. We were recently able to characterize this interaction by considering the energetic barrier to gigaseal formation in terms of colloid theory[4] and now develop the theory to fit data of ion current versus height from deep pushes of cell surfaces. We then show how this understanding allows the quantitative mapping of stiffness across individual cells at low stress, using a variety of cell types in culture – Hippocampal Neuron (HN)[5] cells, a prion protein knockout (*Prnp-/-*) cell line (HpL),[6] and normal, finite lifespan Human Mammary Fibroblast (HMF)[7, 8] cells.

This method is an important technical advance, principally because it allows very soft features of cells to be studied at nanoscale resolution, both in the vertical and lateral directions. To study cells using cantilever techniques, relatively large spheres are typically attached to the tips in order to lower the stress exerted, but this averages out the spatial resolution of differences in stiffness, as well as topography. This same limitation applies to using hydrostatic pressure in ICM as it needs apertures >140 nm[9, 10] in practice (this reduces blockages from unfiltered particulates carried by the flow), and the flow profile is four times wider than this.[9] Fortunately, we find that removing the complication of applying hydrostatic pressure and evaluating the unavoidable forces instead actually makes it more straightforward and less perturbative to image subcellular stiffness, and with higher resolution.

The ability to discern native subcellular structures via stiffness as a second label-free coordinate in addition to topography is itself intrinsically useful, especially given that nanopipets can also deliver reagents to the vicinity[11] and make electrochemical measurements.[12] The detailed knowledge of the structural and mechanical properties in a living cell is just as important though, as these determine the overall mechanical properties of the cell, and their rapid and clearly resolved measurement will further the understanding of how cells respond to forces and changes in their environments.

[a] University Chemical Laboratories, Lensfield Road, Cambridge, CB2 1EW, UK Correspondence: rwc25@cam.ac.uk, dk10012@cam.ac.uk
[b] School of Engineering and Materials Science, Queen Mary University of London, Mile End Road, London, E1 4NS, UK
[c] Centre for Cell Biology and Cutaneous Research, Queen Mary University of London, 4 Newark Street, London, E1 2AT, UK
[d] UCL Institute of Neurology, Queen Square, London, WC1N 3BG, UK

† Electronic supplementary information (ESI) available: Supplementary methods and expanded data presentations of approach curves, stiffness maps, and nanopipet stresses and forces. See DOI: 10.1039/c6sm01106c
‡ These authors contributed equally.

## Results & discussion

### Approach curves

The tip-face of a typical nanopipet usually remains 50-100 nm distant from the plasma membrane during an ICM scan, not approaching closely enough to make contact with cell-surface macromolecules. To minimise the stress on the cell we avoid the complication of applying hydrostatic pressure[9, 13] and ensure the slight weight of the column of solution is balanced by surface tension in the capillary.[14] Thus in these experiments the stress on most cell types is entirely due to the intrinsic colloidal pressure between the cell surface and the glass tip face. This intrinsic pressure σ varies with the size of the gap between the tip face and the plasma membrane as $H/6\pi g^3$ where the Hamaker constant $H$ for the glass-cell interaction across physiological saline is estimated to be 4 zJ [4] and the gap $g$ is found from an empirical model for the drop in ion current as the nanopipet approaches a hard surface, $I = I_0\left(1 - e^{-xg/r}\right)$.

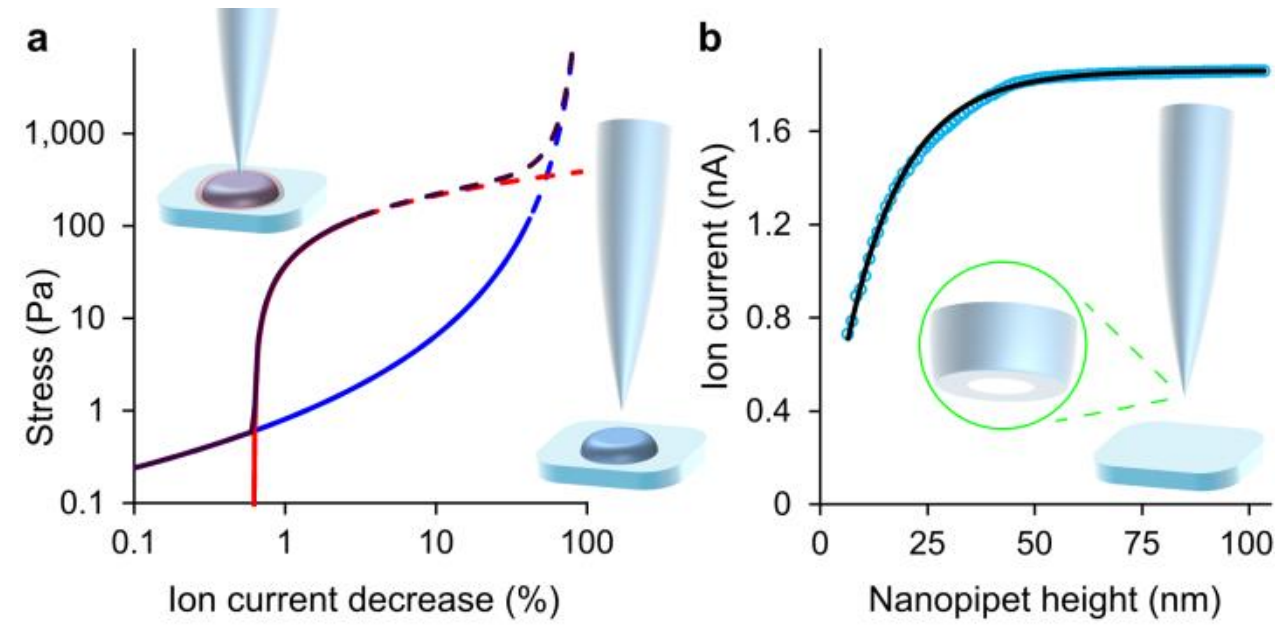

**Fig. 1** At low decreases in ion current the colloidal interaction between the glass tip face of a nanopipet and the cell membrane exerts a miniscule but quantifiable stress. (a) Stress versus ion current decrease calculated for a 100 nm aperture nanopipet. The total stress (purple) is the sum of the intrinsic stress (blue) and, if present, the direct stress on glycocalyx (red), here set to 70.5 nm; 390 Pa.[15] Over 120 Pa (dashes), the cell membrane begins to detach from its anchor-points on the cytoskeleton leading to blebbing.[16] Over 8 kPa the tip-face patches to the membrane.[4] (b) Ion current through an 84 nm aperture nanopipet approaching a hard flat polystyrene surface.

In such approaches, shown in Fig. 1 and Fig. S1 (ESI†), the aperture radius $r$ is determined from the limiting ion current far from the surface via $I_0 = \pi r\kappa V\tan(\alpha)$,[17] where the half-cone angle α is 3 degrees, $\kappa$ is 1.35 Sm$^{-1}$ and $V$ is 200 mV. The tip radius $r$ also determines, along with the empirically determined constant $x = 3.6 \pm 0.2$, the scale of the fall in ion current as the gap narrows. With these values as fixed parameters it is then possible to fit approach data to cells as well, because both ion current and the stresses are fixed functions of the tip-cell gap. For example, the intrinsic stress in terms of the decrease in ion current $\Delta I$ is

$$\sigma = H/6\pi\left((r/x)\ln(I_0/\Delta I)\right)^3 \qquad (1)$$

When combined with the simplest possible models of cell stiffness, this stress already fits approach data to glycocalyx–free cells exactly, as shown in Fig. 2(a and b) and Fig. S2 (ESI†): To fit the first sections of these approaches to neuronal cells just two variables apart from $I_0$ are needed, stiffness and rest height. The height of a cell column of elastic modulus $E$ is $h = h_0\left(1 - (\sigma/E)\right)$, while the height of the tip face above the substrate is $z = h + g$. Writing both $I$ and $z$ parametrically in terms of $g$ then fits approaches to HN cells, on average at $E$ = 93±11 Pa up to the first 0.66±0.09 µm of compression. After pushing this far the current usually begins to decrease faster with height than expected, corresponding to an increase in stiffness. This must correspond to neurons having a stiffer cortex in series with an initially softer range of travel that reaches full compression when conformational slack in the cytoskeleton and in its attachments to the plasma membrane is used up, or when the plasma membrane pushes against the cortex. Thus fitting both sides of the discontinuity in gradient requires a soft portion restricted to non-negative height, with its own stiffness and rest height parameters in series with the cortical parameters. We estimate that for HN cells the cortex is 10.9±0.5 µm at 213±44 Pa plus 0.30±0.05 µm slack at 3.7±0.5 Pa. For HpL cells the cortex is 8.1±0.4 µm at 320±37 Pa plus 0.40±0.03 µm slack at 7.9±1.1 Pa. The different characteristics of slack in HpL cells may be related to absence of PrP or to ectopic expression of Dpl,[18] and hence to ataxia in HpL mice.[18]

Having detected the plasma membrane at low stress it is remarkable that if our measurements had not pushed far enough to use up the slack there would have been no indication of its existence, for in the initial regime of compression the dual stiffness fit is identical to the uniform model. Its stiffness and rest height parameters for the cortex and slack even combine analytically in the following simple formula to give exactly the same apparent stiffness:

$$E = (h_C + h_S)/\left((h_C/E_C) + (h_S/E_S)\right) \qquad (2)$$

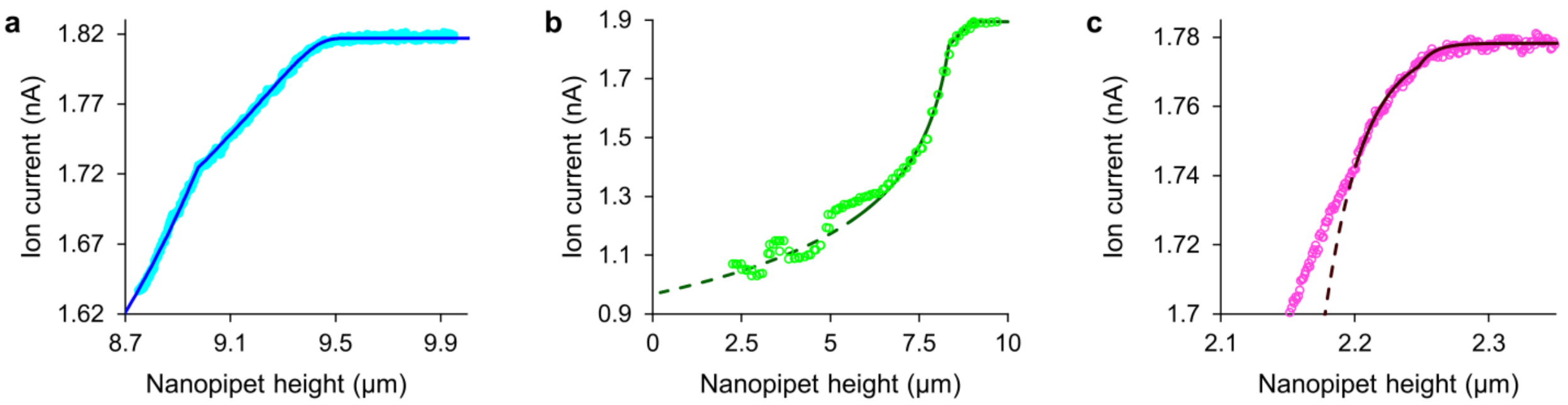

**Fig. 2** With stress characterized in terms of ion current decrease, approach data fits simple models of cell mechanics: (a) HN cell, apparent stiffness 107 Pa for 0.48 µm, fits 260 Pa, 9.14 µm cortex with 5.5 Pa, 0.29 µm slack. (b) HpL cell, apparent stiffness 52 Pa for 0.70 µm, fits 350 Pa, 8.4 µm cortex with 4 Pa, 0.6 µm slack. (c) HMF cell fits 9.0 kPa, 2.2 µm height with 62 nm glycocalyx. The fit lines are shown dashed at the membrane detachment stress of 120 Pa.

This initial slack in neurons is interesting as it would account for observed changes in the volume of the brain's interstitial space[19] if there were to be a slight rise in interstitial fluid pressure during the transition to sleep. A distributed pressure differential like this could arise osmotically, or from upregulation of astrocytic AQP4 aquaporins, which would lower their resistance to cerebrospinal fluid pressure[20] and arterial pulsation.[21] The absence of active cellular contraction in this mechanism avoids opposing forces that would break synapses, while the gentle compression of each neuron would allow it to efficiently expunge waste metabolites and misfolded proteins through cellular pores, to be washed away by glymphatic flow.

Some other cell types are coated by a porous network of proteoglycans called the glycocalyx. The tip face compresses this elastically when in contact, allowing the stiffness of such cells also to be determined without close approach to the plasma membrane. Any glycocalyx is only strained when the gap $g$ is less than its thickness $t$, generating a direct stress $\varsigma = Y\left(1 - (g/t)\right)$ that adds to the intrinsic pressure on the membrane as shown in Fig. 1. In this paper we take the elastic modulus of HMF glycocalyx $Y$ to be that of human umbilical vein endothelial cell glycocalyx, 390 Pa.[15] The decrease in ion current with height is then fully determined by the elastic modulus of the cell and the thickness of the glycocalyx around it. Conversely, these parameters can be inferred in order to fit data of ion current versus tip height, as shown in Fig. 2(c) and Fig. S2 (ESI†). Our HMF approach data fit means of 3.9±0.6 kPa cell stiffness and 70.5±1.6 nm glycocalyx thickness, comparable to the 39.5 nm thickness determined for erythrocytes[22] and 3.2 – 75.0 nm for endothelial cells.[23]

Although the intrinsic stress increases sharply with $\Delta I$, it fully compresses most cells well before reaching its 8 kPa maximum when, in the absence of glycocalyx, the tip face seals to the membrane.[3, 4] Thus soft cells like neurons cannot be patched without applying negative hydrostatic pressure, and ion current during approaches is rarely asymptotic to zero. At full compression, where the apical and basal membranes are pushed together against the substrate, sealing would be quickly followed by membrane rupture, at 3 MPa.[24] We did not push the cells this far. A much earlier consideration when compressing a cell is the piece-wise detachment of the plasma membrane from its anchor-points on the cytoskeleton, which begins at intracellular pressures of 45-300 Pa.[16] This phenomenon allows a cell to accommodate distortion without bursting, and is known as blebbing. Cells actively re-attach folds of plasma membrane to the cytoskeleton, so a live cell is able to wrinkle blebs back into place. However, not exceeding the blebbing stress in the first place maintains a passive elastic response that does not require this energy expenditure. A good estimate of the stress at which these effects typically begin is the log-mean of the above range, 120 Pa. Around this point, where the lines in all figures become dashes, the ion current can decrease slower than expected due to gradual membrane detachment decreasing the effective stiffness, thereby maintaining the tip-cell gap. Any sudden blebbing can reduce the intracellular pressure to such an extent that the gap actually re-widens, whereupon the ion current jumps upwards, as in Fig. 2(b) and S2 (ESI†). The ion current sometimes decreases faster than expected instead, indicating a second increase in cortical stiffness.

## Low stress mapping

In contrast to the approach data which extend to large decreases in ion current, to map cell stiffness we determine the height of each point above the substrate at two low set decreases in ion current, typically only 0.3% and 1.5%. Two fields of nanopipet heights are thus measured at two constant stresses; at a minimal stress of 0.1-10 Pa and at a compressive stress of 1-100 Pa, where both are precisely determined in any particular scan by its specific parameters, described earlier. The nanopipet heights are converted to cell heights by subtracting the tip-cell gap, which typically decreases from 50 to 30 nm for neuronal cells (or from 80 to 50 nm for fibroblasts, where the pipet must be wide enough to detect the cell surface before pushing into the glycocalyx). The stresses and corresponding cell heights are thereby accurately evaluated even though the imaging process only momentarily pushes the cell surface around 100 nm on average at each point. The two simultaneous equations $E = \Sigma_1/\varepsilon_1 = \Sigma_2/\varepsilon_2$, where $\Sigma$ is the total stress $\sigma + \varsigma$ and $\varepsilon$ denotes strain, then give the cell stiffness as:

$$\begin{aligned} E &= ((\Sigma_2 - \Sigma_1)h_1/(h_1 - h_2)) + \Sigma_1 \\ &= (\Delta\Sigma(z_1 - g_1)/(\Delta z - \Delta g)) + \Sigma_1 \qquad (3) \end{aligned}$$

Taking account of the changes in separation and stress in this way allows the nanopipet to discern differences in stiffness across individual cells, for example over actin stress fibres and apparent endocytotic events that are not visible in the topographies, as shown in Fig. 3, 4 and Fig. S3-S7 (ESI†).

Note that the stiffness maps of fibroblasts need an independent estimate of glycocalyx stiffness and also its thickness as determined from approach data. Glycocalyx stiffness does depends on cell type; it is 250 Pa for pulmonary endothelial cells for example,[25] but the dominant direct stress is linear in this parameter so changing it does not affect the contrast in stiffness that will usually be of primary interest.

To compare our measurements with other techniques that have assayed cell body stiffness we calculated the mean stiffness of somatal regions in our scans as identified by topography, demonstrated in Fig. S3 (ESI†). The hippocampal neurons have an apparent cell body stiffness of 56±9 Pa, corresponding to a cortical stiffness of 310±109 Pa. When assessed by 6 μm diameter polystyrene spheres on a cantilever tip the stiffness of similar cell bodies was reported to be 900 Pa,[26] suggesting the cantilever spring constant was too high to detect the initial slack and instead measured the cortical stiffness directly. Some HN cells we measured did have a cortex this stiff but others were much softer; we suspect that in earlier cantilever studies this softer subpopulation would have been flattened against the substrate and missed. HpL cells had an apparent cell body stiffness of 64±4 Pa, corresponding to a cortical stiffness of 702±22 Pa. The stiffness we find for HMF cells, 2.25±0.27 kPa, is comparable to an average stiffness from force microscopy of fibroblasts,[27, 28] 2.89±0.28 kPa. These cells are stiffened by the enhanced lateral force transmission of the numerous stress fibres[29] seen in Fig. 3(e) and 4(e).

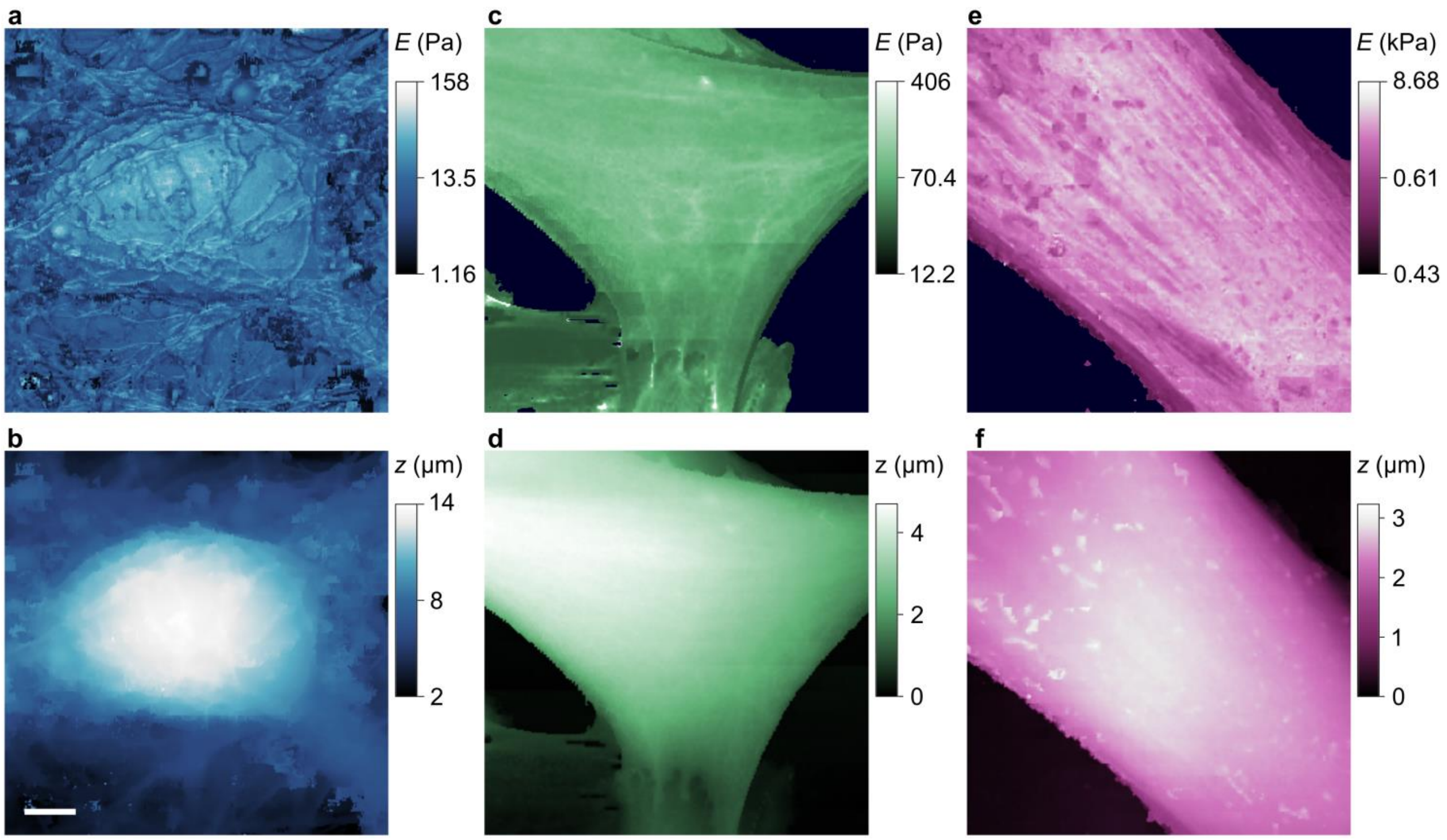

**Fig. 3** Stiffness and topography of three cell types imaged at high resolution by nanopipet ICM. (a, b) HN neuron mapped by a 110 nm aperture nanopipet at Δ*I*=0.3%, 2% exerting stress of 0.29 Pa, 0.96 Pa at tip-cell gaps of 90 nm, 60 nm that typically push the cell soma 0.1-0.2 µm. (c, d) HpL cell mapped by a 52 nm aperture nanopipet at Δ*I*=0.6%, 3% exerting stress of 4.27 Pa, 13.26 Pa at gaps of 37 nm, 25 nm. (e, f) HMF fibroblast mapped by a 100 nm aperture nanopipet at Δ*I*=0.3%, 3% exerting stress of 0.4 Pa, 122.5 Pa at gaps of 81 nm, 49 nm, resolving the stiffness of stress fibres. The substrate stiffness is masked out in dark blue. The lateral scale bar is 4 µm.

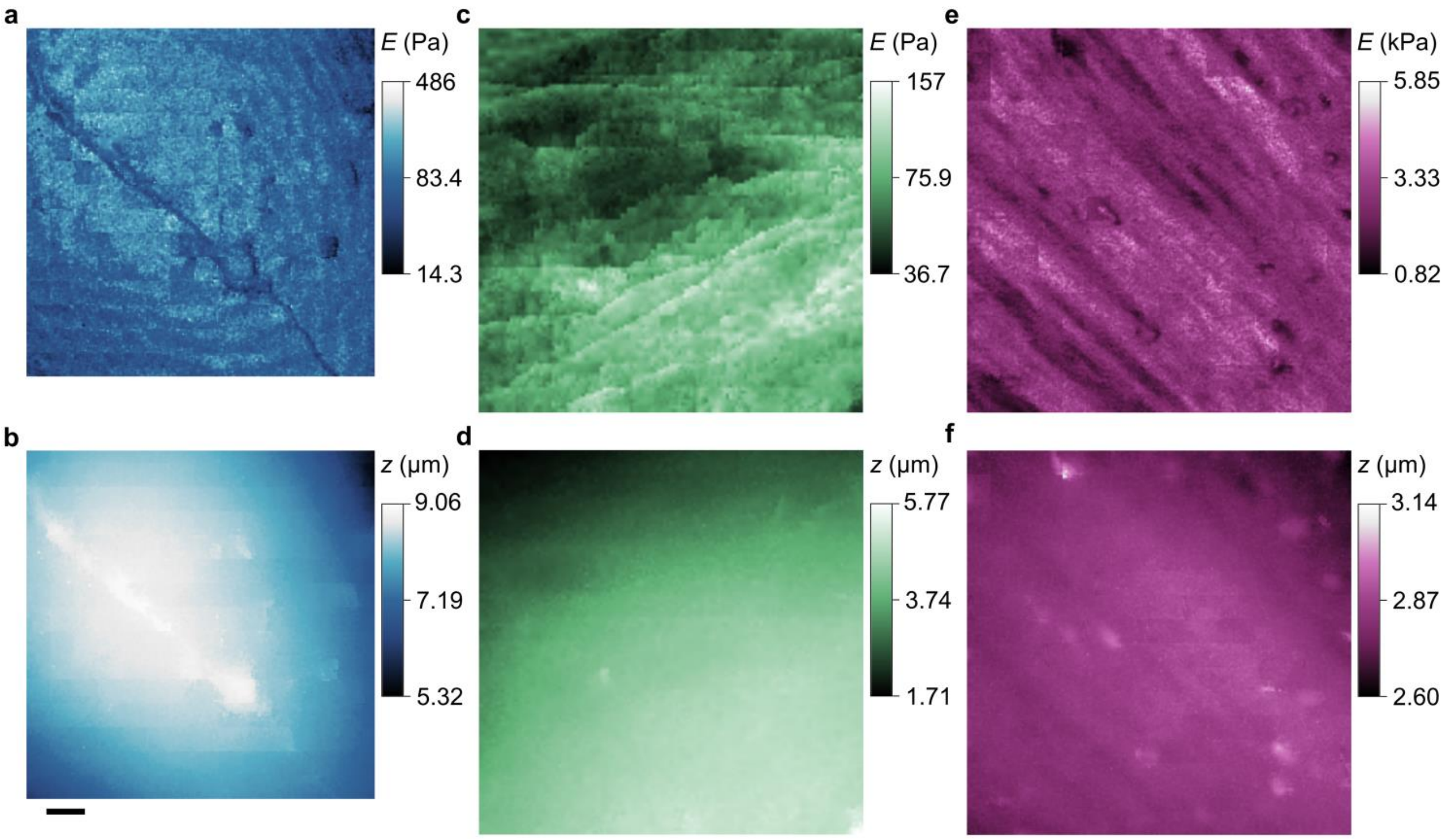

**Fig. 4** Stiffness and topography of three cell types imaged at high resolution by nanopipet ICM. (a, b) HN soma mapped by a 62 nm aperture nanopipet at ΔI=0.5%, 1%, exerting stress of 2.23 Pa, 3.40 Pa at gaps of 46 nm, 40 nm. The lateral scale bar is 1 µm. (c, d) HpL soma mapped by a 52 nm aperture nanopipet at Δ*I*=0.6%, 4.2%, exerting stress of 4.27 Pa, 17.95 Pa at gaps of 37 nm, 23 nm. (e, f) Apical area of HMF fibroblast mapped by a 100 nm aperture nanopipet at Δ*I*=0.3%, 3%, exerting stress of 0.4 Pa, 122.5 Pa at gaps of 81 nm, 49 nm. Cell surface structures, probably endocytotic events, are visible in the stiffness map but not in the topography.

As a reference for future studies, the dependencies and limiting factors of the stresses in ICM are illustrated in Fig. S8 (ESI†). Each curve begins at the minimum detectable ion current decrease in 1 ms, calculated for a signal to noise ratio of three times the thermal noise, $\Delta I_{RMS} = \sqrt{4k_B T \cdot \Delta f / R}$.[30] These values increase for smaller aperture diameters but it would be possible to detect smaller changes in ion current for higher resistance nanopipets by extending the acquisition time. The bandwidth $\Delta f$ = 1 kHz corresponds to the rate of data acquisition typically necessary for imaging experiments.

## Conclusions

These are general methods for assaying and imaging cell stiffness but they have already identified here some specific features of interest. We have shown how to determine the thickness of the glycocalyx from approach data, and have found that some subcellular structures exhibit strong contrast in stiffness but none in topography. We have also identified that the initial deformation of most points of the plasma membrane of neurons is extremely soft, indicating that we must often be encountering the spaces between its non-tethered points and the cortical cytoskeleton, and/or conformational slack in the cytoskeleton itself.

Overall, these results demonstrate that it is possible to map the stiffness of cells at very high resolution, both laterally and vertically, without the considerable effort of modifying ICM apparatus to apply hydrostatic pressure. The absence of flow also allows narrower nanopipets to be used that would otherwise be prone to blockages, and for which the forces we describe would have to be evaluated in any case. Further advantages stem from minimizing the offset from the tip-face to the cell surface – if applying hydrostatic pressure this offset is necessarily larger to accommodate the flow profile, which lowers resolution and begins pushing the cell before its surface is detected. The equations developed here also indicate that it may be possible to patch hard cells without the requirement of applying negative hydrostatic pressure.

Besides its ability to map a vast range of stiffness at the nanoscale, stress-quantitative ICM will now enable many other interesting studies of live cells, including fundamentally non-invasive assays of differentiation, subcellular response, and mechanosensation. It will be possible for example to assay the exact stresses at which mechanosensitive ion channels open. Thus this advance in understanding of ICM greatly increases its versatility for nanoscale biophysics and the study of cellular mechanics.

## Author contributions

R.W.C. developed the models and experimental design; P.N. and R.W.C. designed the apparatus and scanned the cells; P.N., R.W.C. and A.Z. coded the controllers; P.N., A.Z. and O.R. recorded the approaches; R.W.C., P.N. and A.Z. processed the data; M.C.-J. and K.V. contributed the HN cells; A.D. cultured the HpL cells; E.J.T. and C.B. contributed the HMF cultures; R.W.C. and D.K. directed the research.

## Acknowledgements

We thank Prof. P. St George-Hyslop and Prof. Dr A. Aguzzi for the HpL cells, Dr M. Stampfer for the HMF cells, and Dr S. Antolin for comments. This work was supported by the BBSRC (BB/L006227/1), EPSRC (EP/H01098X/1) and MRC (G0701057, MR/K501372/1). A.D. was funded by a Herchel Smith Postdoctoral Fellowship and O.R. by the RSC Analytical Chemistry Trust Fund.

Table of contents entry:

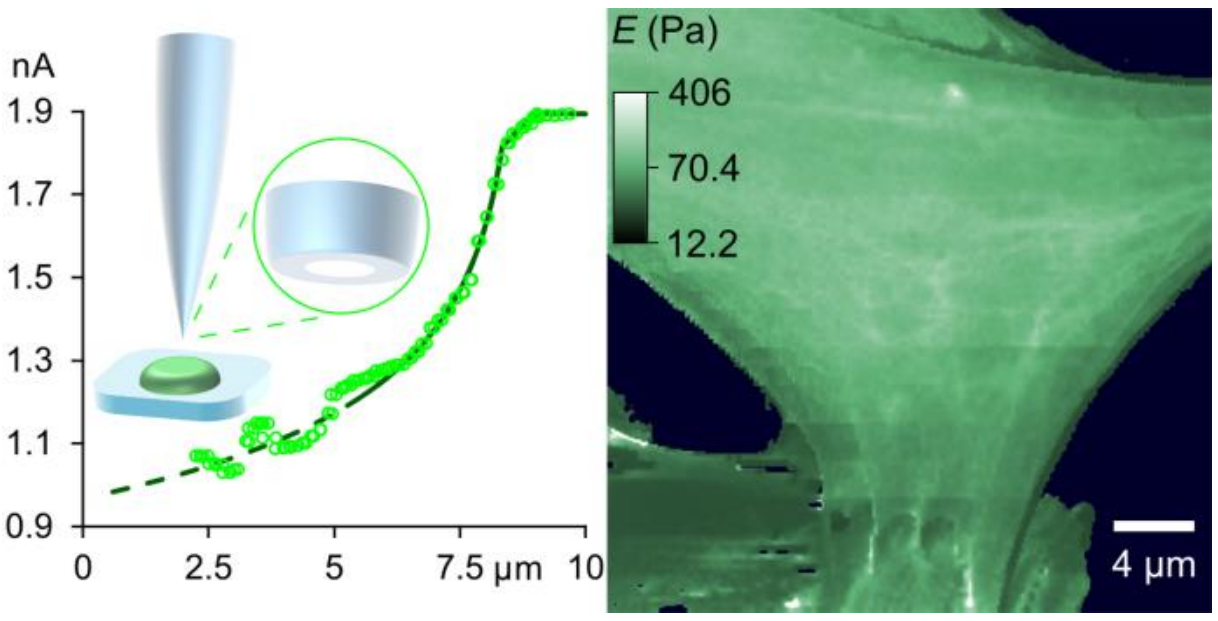

*Quantifying forces inherent to Ion Conductance Microscopy enables it to map the stiffness of sub-cellular structures, even if very soft.*

 

**Endnote References**

Electronic Supplementary Information

# Low Stress Ion Conductance Microscopy of Sub-Cellular Stiffness

## Electronic Supplementary Information

Richard W. Clarke,‡*[a] Pavel Novak,‡[b] Alexander Zhukov,[a] Eleanor J. Tyler,[c] Marife Cano-Jaimez,[d] Anna Drews,[a] Owen Richards,[a] Kirill Volynski,[d] Cleo Bishop[c] and David Klenerman*[a]

## CONTENTS

[e.] University Chemical Laboratories, Lensfield Road, Cambridge, CB2 1EW.
[f.] School of Engineering and Materials Science, Queen Mary University of London, Mile End Road, London, E1 4NS.
[g.] Centre for Cell Biology and Cutaneous Research, Queen Mary University of London, 4 Newark Street, London, E1 2AT.
[h.] UCL Institute of Neurology, Queen Square, London, WC1N 3BG.
‡ These authors contributed equally.
* Correspondence to: rwc25@cam.ac.uk; dk10012@cam.ac.uk.

# Electronic Supplementary Information

## Supplementary Methods

### Apparatus

We used ICM instruments based in London, described previously,[1] and Cambridge. The Cambridge instrument, based on an Eclipse Ti-U inverted microscope (Nikon), was built on a MK26 vibration isolation table (Minus k) and enclosed inside a Faraday cage. The stage was stabilized by a kinematic mount from an aluminium top-plate to a 12 mm M-461-XY-M stage (Newport) fixed to an aluminium underplate and adjusted by micrometers to align the optical axis. Samples are positioned by two DM-13L lockable differential micrometers (Newport) mounted horizontally on the underplate at 45 degrees to the microscope eyepiece, tensioned by steel tensators to hardened steel stops on an aluminium block holding a 30 µm XY P-733.2DD piezoelectric drive (PI) with a plastic push-in dish mount for 35 mm dishes (Corning). Nanopipets are clamped vertically into a v-groove on an acrylic pipette holder on a 25 µm P-753 piezoelectric drive (PI) mounted below the current amplifier headstage (Axopatch 200B, Molecular Devices), both held to a 25mm M-462-X-M crossed-roller bearing translation stage (Newport) driven by a lead-screw 25mm M-112.1DG motor (PI), kept to one extreme of yaw by a tensator. Pairs of nanopipets were made by a P-2000 puller (Sutter), pulling 10 cm long, 0.5/1.0 mm inner/outer diameter fire-polished borosilicate glass capillaries with filaments (Intracel). With the program Heat 350, Fil 3, Vel 30, Del 220, Pul 0, Heat 390, Fil 3, Vel 40, Del 180, Pul 255 we made small, 400-200 MΩ, nanopipets. Half of these were already in the target range of 300-100 MΩ; a 50% increase in current was made as needed by flattening the tip face against the sample dish at 100 $nm.ms^{-1}$ whilst ensuring the nanopipet rapidly rose at the first sign of any current decrease. Larger, typically 125-80 MΩ nanopipets were made with the program Heat 310, Fil 3, Vel 30, Del 160, Pul 0, Heat 330, Fil 3, Vel 25, Del 160, Pul 200. All experiments applied 200 mV and used physiological solutions, of conductivity 1.35 $S.m^{-1}$. Actuators were commanded by customized hopping mode[2, 3] software on digital controllers (Ionscope). The scanning control sequence ascertains the ion current in bulk solution while shifting the horizontal sample position, then lowers the nanopipet to register the heights at which the ion current decreases by set fractions, typically 0.3% and 2%, before repeating the cycle. The approach control sequence records ion current at many heights as the nanopipet is lowered. Approaches to HpL cells also alternated the changes in height between increases and decreases, verifying that the cells responded elastically in the ranges assayed.

### Computation

All curves are computed parametrically in terms of the tip-surface gap and the graphs of intrinsic and direct stress show the full range – many cell types would be fully compressed at lower total stresses than these curves reach. The force curves are calculated by multiplying the stress by the area of the tip face, evaluated from the conserved ratio of wall thickness to inner diameter in nanopipets pulled from glass capillaries. The parametric equations for fitting approach curves to cells described in the main text may be summarised in terms of the functions defined there as $[z, I] = \left[h\left(\sigma(g) + \varsigma(g)\right) + g, I(g)\right]$ in general and for neuronal cells as $[z, I] = \left[h_{(C)}\left(\sigma(g)\right) + h_{(S)}\left(\sigma(g)\right) + g, I(g)\right]$. Images of topographic and stiffness maps were rendered using the Gwyddion scanning probe microscopy software suite.[4] The fits in Figs. 2 and S2 are representative of 16, 37 and 11 approaches for HN, HpL, and HMF cells respectively. Fourteen of the sixteen HN approaches required the dual stiffness model; the other two must have assayed points on the cell surface that were directly attached to the cytoskeleton. To avoid singularities in stiffness maps including substrate we limit gap-corrected height difference fields to a non-zero minimum of 1 nm. The HN maps in Figs. 3, S3 & S4 are representative of eleven scans, two of which were non-independent zooms not repeated in the averages quoted in the main text. The highest parts of these cells were pushed down by around 200 nm. The stiffness maps of HpL cells in Figs. 3 and S5 are a special case where the pairs of height measurements were not acquired during a single scan but were instead determined from two consecutive scans. We linearly aligned these pairs of topographies by reference to the substrate which in these circumstances already accounts for the difference in pipette-cell gap and slight thermal drift. However, in between the two scans for Fig. 3(c) the cell actively moved upwards in a few places, creating point artifacts of high stiffness. In order to prevent these from impairing contrast we have limited the stiffness scale of this particular map to three times the root mean square value. This issue does not affect any of the other stiffness maps, whose pairs of height measurements were acquired in rapid succession at each point of single scans, an improved method. The two HpL maps in Fig. S5 are independent scans; the absolute height of the substrate in the right-hand panel was determined by reference to an earlier scan. The HMF maps in Figs. S6 & S7 are representative of seven scans, three of which were non-independent zooms not repeated in the averages quoted in the main text. Due to noise in the ion current some scans have two or three outlying height measurements that rise more than 1 µm above the surrounding topography. To improve contrast it is routine in ICM imaging to replace these spikes with averages of the surrounding heights, a minor correction as illustrated by the similarity of Figs. 3(f) and S6. For the purpose of illustration the latter topography is left uncorrected for the effect of active cell movement during the scan, which creates mismatches between the bands of measurement points. Where discernible, we minimize these topographic mismatches by piece-wise linear regression of the stripe differences, retaining the original fields of measured height differences for the stiffness map calculations.

**Materials**

All chemicals were reagent grade and used as supplied. Solutions were filtered by Anotop 0.02 µm filters (Whatman). All measurements were performed in polystyrene dishes (Corning), unless otherwise noted.

**Cells**

Hippocampal neurons: HN cells were cultured in Neurobasal A/B27 based medium on an astrocyte feeder layer plated on 19 mm glass coverslips covered with poly-D-lysine[5] having been isolated from humanely killed P0–P2 rats.[6] The standard extracellular solution used in experiments on hippocampal neurons contained 125 mM NaCl, 2.5 mM KCl, 2 mM $MgCl_2$, 2 mM $CaCl_2$, 30 mM glucose, 0.01 mM NBQX, 0.05 mM APV, and 25 mM HEPES (pH 7.4). HpL neuronal cells: The HpL cell line is derived from mouse hippocampal neurons transformed by SV40 [7] and is also known as P4. These cells were grown in 75 $cm^2$ flasks with OptiMEM (Invitrogen) plus 10% fetal bovine serum, 0.5% penicillin, 0.5% streptomycin in humid air at 37˚C, 5% $CO_2$. For ICM experiments the cells were cultured to 70–90% confluence in 35 mm culture dishes (Corning) with 14 mm glass microwell, number 1 thickness cover glass (MatTek). Scans were conducted in Leibovitz-15 (L15) medium (Invitrogen), 1-2 days after plating. Mammary fibroblasts: Normal human mammary fibroblasts were cultured as previously described[8] with L15 medium as the bath solution for ICM experiments.

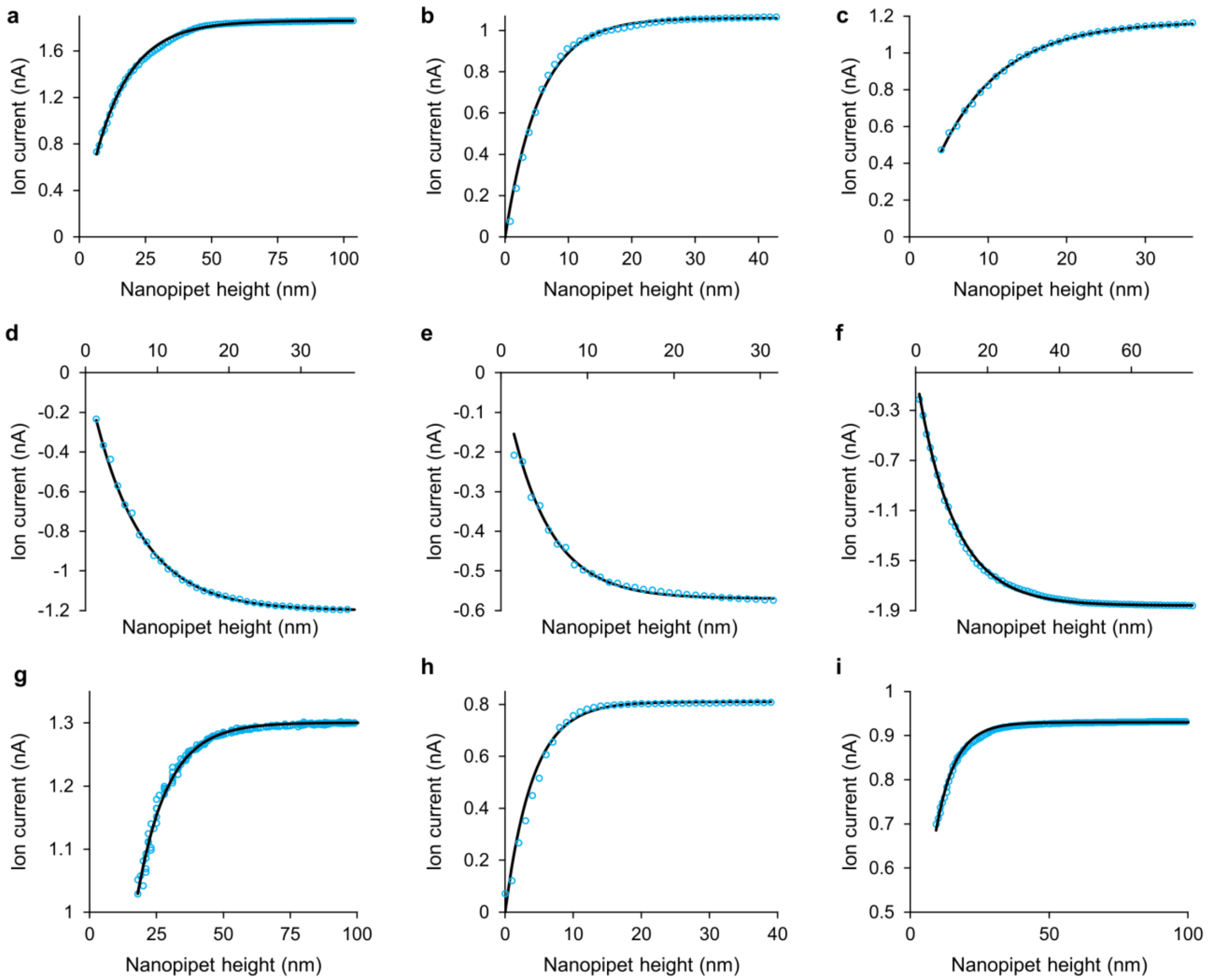

Figure S1. Approach calibrations. (**a**) 84 nm aperture (108 MΩ) nanopipet, fitting x=3.1. (**b**) 48 nm aperture (189 MΩ) nanopipet, fitting x=4.4. (**c**) 53 nm aperture (171 MΩ) nanopipet, fitting x=3.3. (**d**) 54 nm aperture (167 MΩ) nanopipet, fitting x=4. (**e**) 26 nm aperture (351 MΩ) nanopipet, fitting x=2.7. (**f**) 84 nm aperture (108 MΩ) nanopipet, fitting x=4. (**g**) 59 nm aperture (154 MΩ) nanopipet, fitting x=2.55. (**h**) 36 nm aperture (247 MΩ) nanopipet, fitting x=4.5. (**i**) 42 nm aperture (215 MΩ) nanopipet, fitting x=3. Mean geometric factor x=3.6±0.2 (N=19).

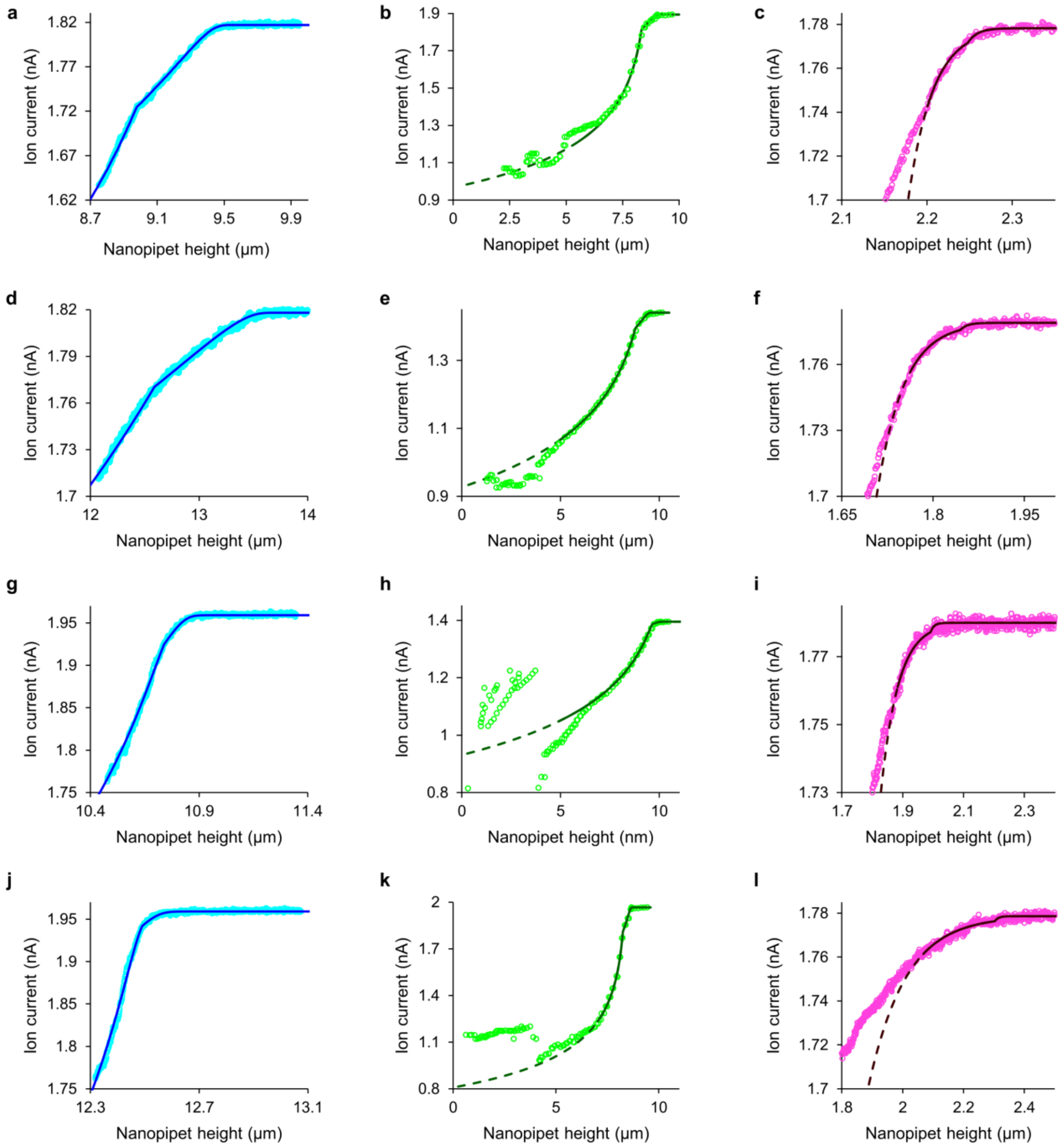

Figure S2. Cell approaches. (**a**) HN, apparent stiffness 107 Pa for 483 nm, fits 260 Pa cortex 9.135 μm thick with 5.5 Pa, 290 nm slack. (**b**) HpL, apparent stiffness 51.7 Pa for 0.7 μm, fits 350 Pa, 8.4 μm cortex with 4 Pa, 0.6 μm slack. (**c**) HMF fits 9.0 kPa, 2.2 μm with 62 nm glycocalyx. (**d**) HN, apparently 40.4 Pa for 1.01 μm, fits 81 Pa, 13.03 μm cortex with 3 Pa, 525 nm slack. (**e**) HpL, apparently 90.1 Pa for 738 nm, fits 265 Pa, 9 μm cortex with 7 Pa, 0.5 μm slack. (**f**) HMF fits 3.5 kPa, 1.775 μm with 70 nm glycocalyx. (**g**) HN, apparently 171 Pa for 111 nm, fits 337 Pa, 10.75 μm cortex with 1.75 Pa, 55 nm slack. (**h**) HpL, apparently 50.4 Pa for 0.5 μm, fits 240 Pa, 9.7 μm cortex with 2.5 Pa, 0.4 μm slack. (**i**) HMF, 2.6 kPa, 1.92 μm with 72 nm glycocalyx. (**j**) HN, apparently 175 Pa for 78 nm, fits 760 Pa, 12.45 μm cortex with 1.09 Pa, 60 nm slack. (**k**) HpL, apparently 160 Pa for 375 nm, fits 770 Pa, 8.3 μm cortex with 7 Pa, 0.3 μm slack. (**l**) HMF, 1.25 kPa, 2.23 μm with 74 nm glycocalyx.

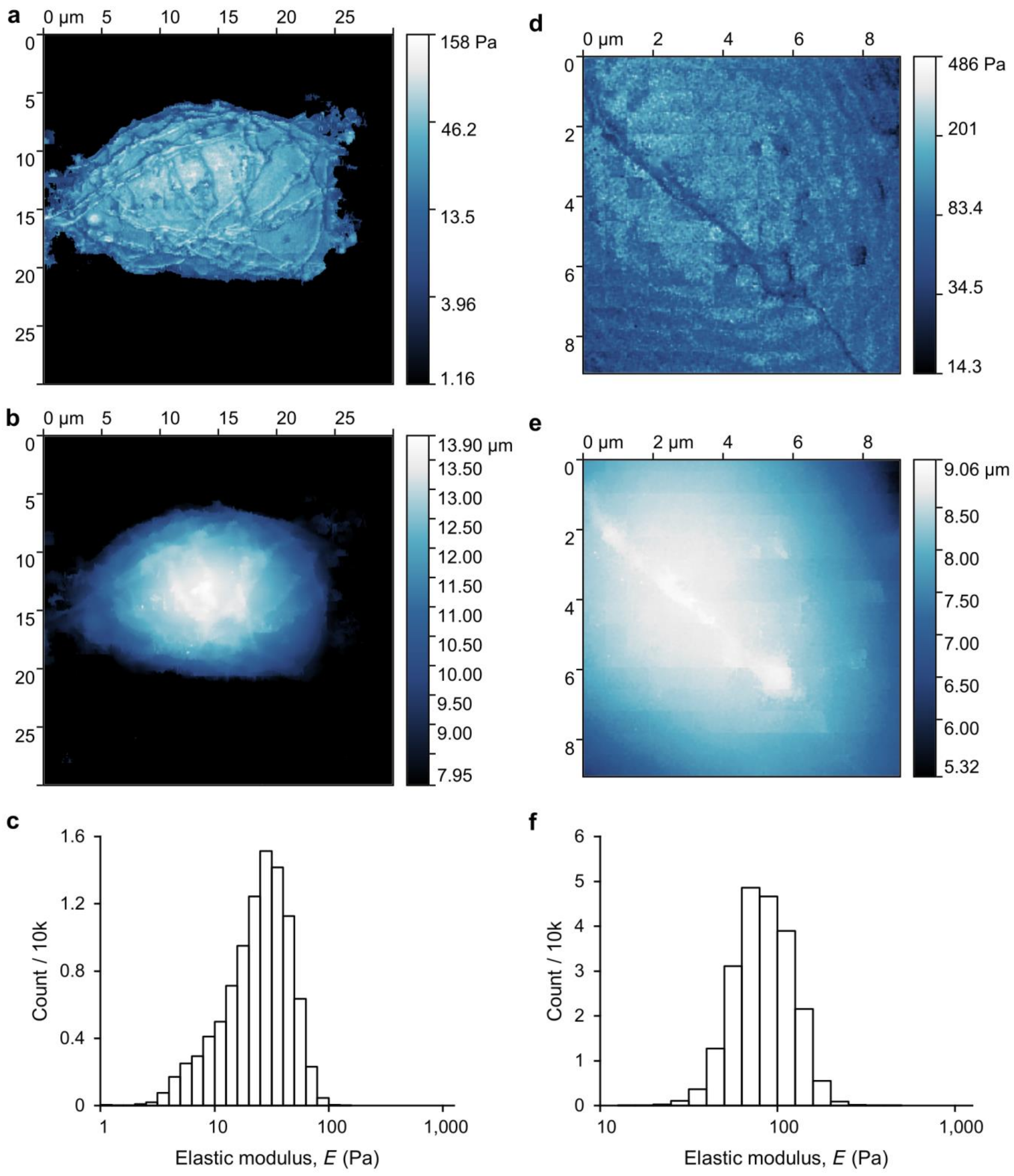

Figure S3. Maps of hippocampal neuron cell stiffness and topography. (**a**) Cell body stiffness, selected as the top 50% topography, mapped with an 81 MΩ nanopipet of 110 nm aperture at $\Delta I$=0.3%, 2%, exerting stress of 0.29 Pa, 0.96 Pa at gaps of 90 nm, 60 nm respectively. (**b**) Top 50% topography. (**c**) Histogram of cell body stiffness, mean 28 Pa. (**d**) Higher resolution map of the top of the cell body of a hippocampal neuron, 9.063 µm by 9.063 µm, mapped with a 145 MΩ nanopipet of 62 nm aperture at $\Delta I$=0.5%, 1%, exerting stress of 2.23 Pa, 3.40 Pa at gaps of 46 nm, 40 nm respectively. (**e**) Topography. (**f**) Stiffness histogram, mean 89 Pa.

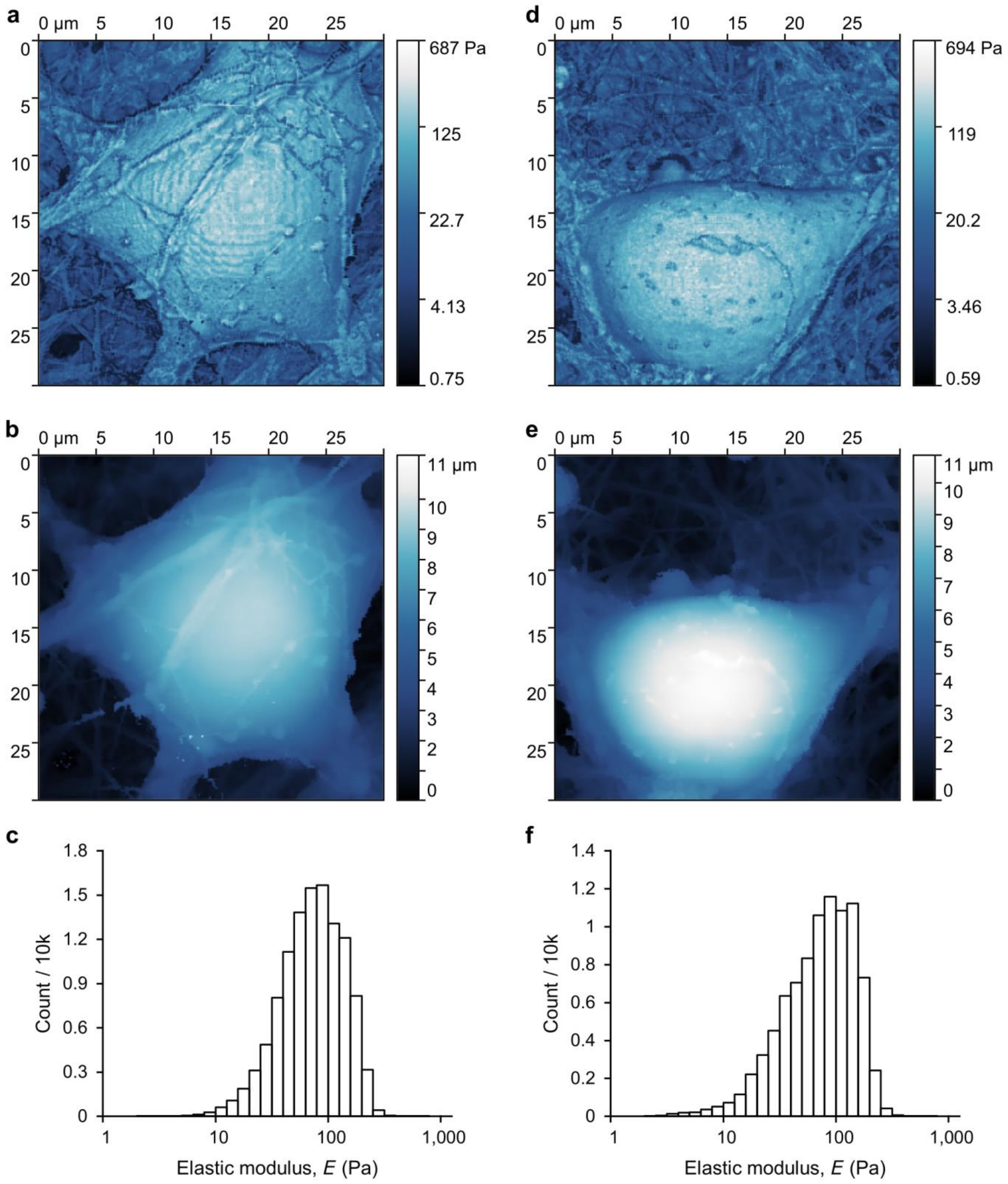

Figure S4. Maps of hippocampal neuron cell stiffness and topography. (**a**) Cell stiffness mapped with a 110 MΩ nanopipet of 81.8 nm aperture diameter at $\Delta I$=0.3%, 1%, exerting stress of 0.74 Pa, 1.48 Pa at gaps of 66 nm, 52 nm respectively. (**b**) Topography. (**c**) Histogram of cell body stiffness, selected as top 50% topography; mean 86.6 Pa. (**d**) Cell stiffness mapped with a 102 MΩ nanopipet of 88.2 nm aperture diameter at $\Delta I$=0.3%, 1.5%, exerting stress of 0.59 Pa, 1.56 Pa at gaps of 71 nm and 51 nm respectively. (**e**) Topography. (**f**) Cell body stiffness, selected as top 50% topography; mean 87.5 Pa.

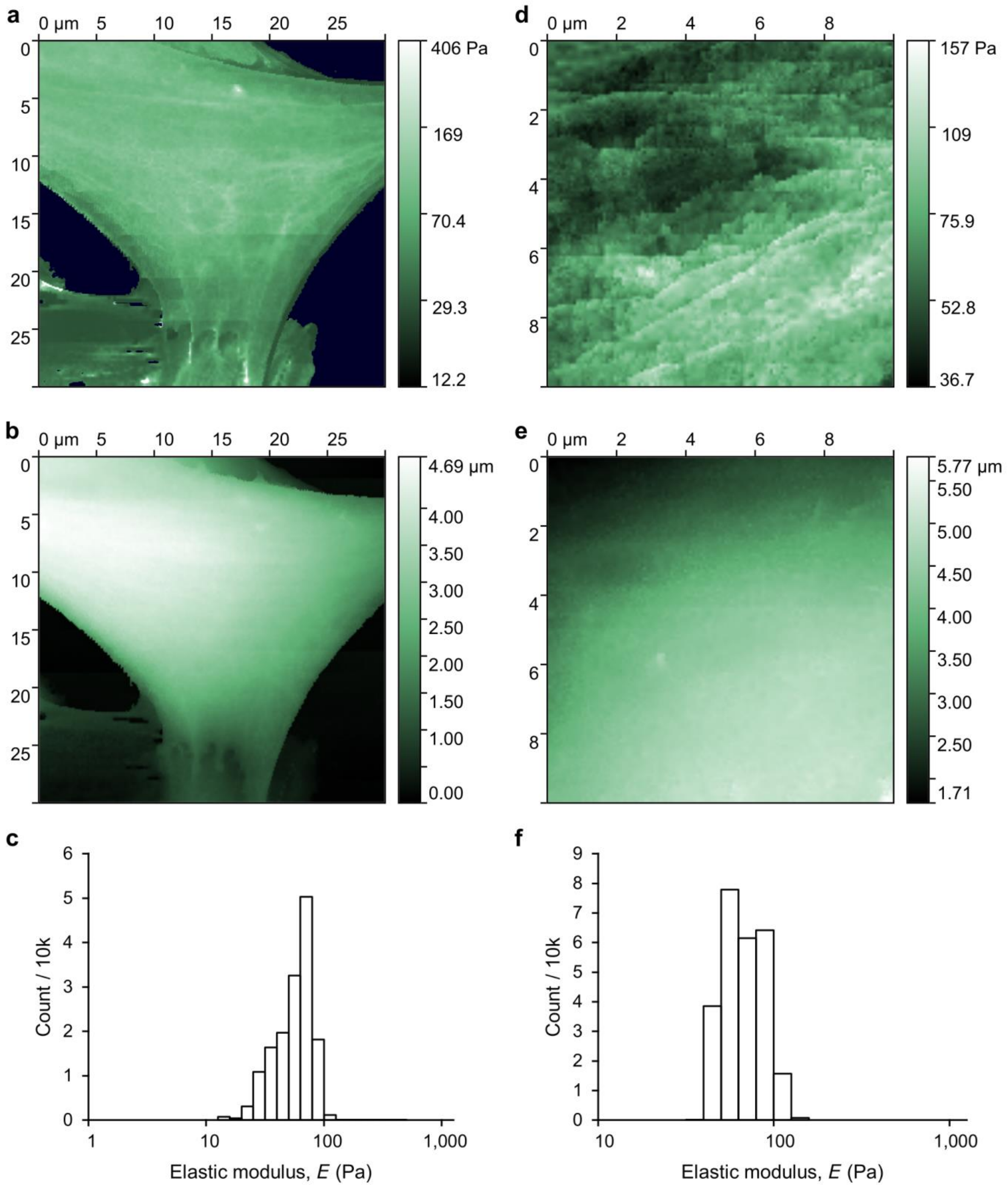

Figure S5. Maps of HpL neuronal cell stiffness and topography. (**a**) Cell stiffness mapped with a 174 MΩ nanopipet of 51.8 nm aperture diameter at $\Delta I$=0.6%, 3%. This exerts stress of 4.27 Pa, 13.26 Pa at gaps of 37 nm, 25 nm. (**b**) Topography. (**c**) Histogram of cell body stiffness, selected as top 70% topography; mean 58.7 Pa. (**d**) Higher resolution map of the stiffness of a 10 µm by 10 µm area on top of a different cell mapped with the same nanopipet at $\Delta I$=0.6%, 4.2%, exerting stress of 4.27 Pa, 17.95 Pa at gaps of 37 nm, 23 nm. (**e**) Topography. (**f**) Stiffness histogram, mean 70 Pa.

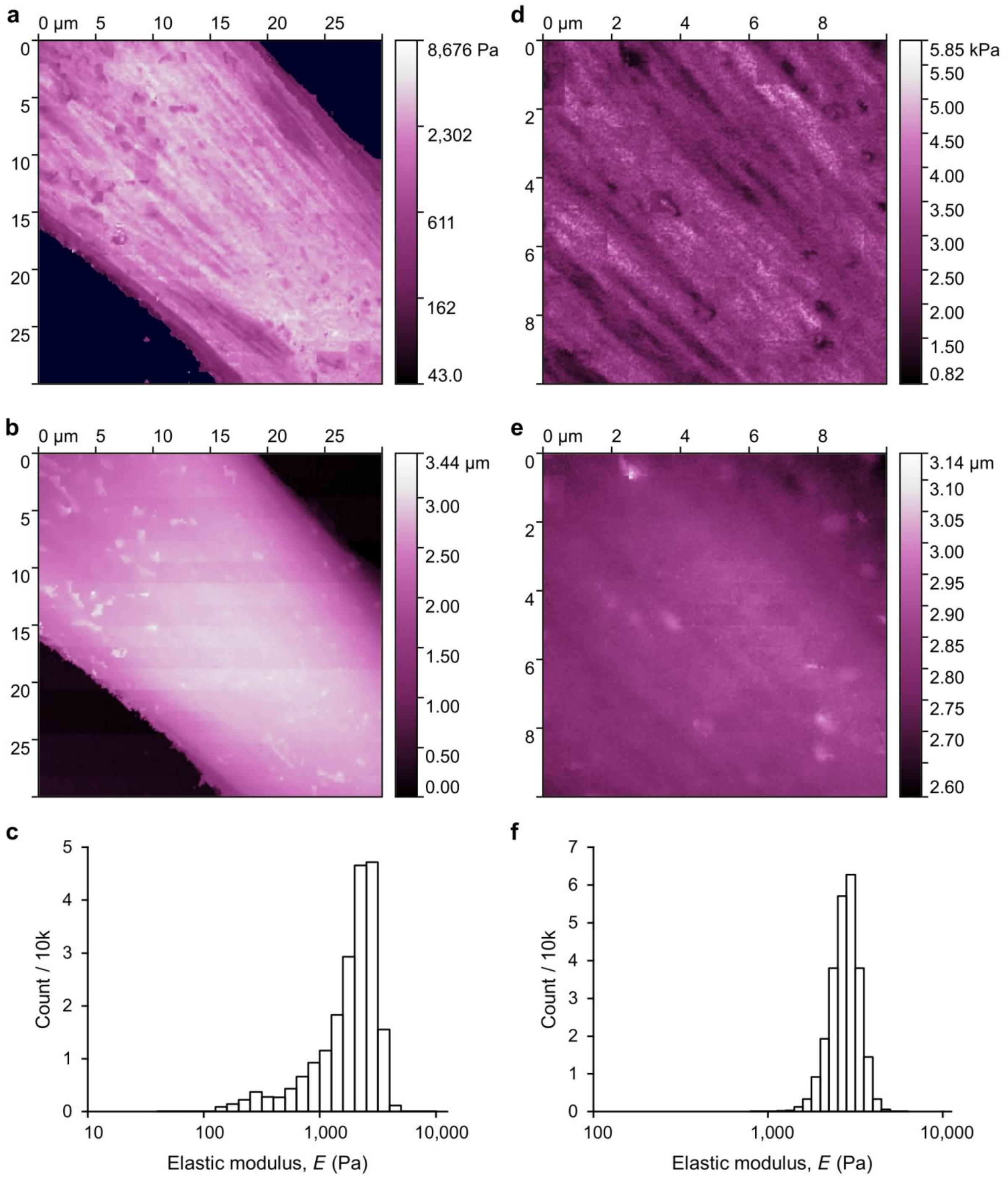

Figure S6. Maps of mammary fibroblast cell stiffness and topography. (**a**) Cell stiffness mapped with a 90 MΩ nanopipet of 100 nm aperture diameter at $\Delta I$=0.3%, 3.0%, exerting stress of 0.4 Pa, 122.5 Pa at gaps of 81 nm, 49 nm. (**b**) Topography. For comparison this topography is not corrected for active cell movement during the data acquisition as detailed in methods. (**c**) Histogram of cell stiffness, selected by top 90% topography; mean 1.992 kPa, mode 2.506 kPa. (**d**) Higher resolution map of the stiffness of a 10 μm by 10 μm area on the top of the same cell using the same stress levels. (**e**) Topography. (**f**) Histogram of stiffness measurements, mean 2.792 kPa.

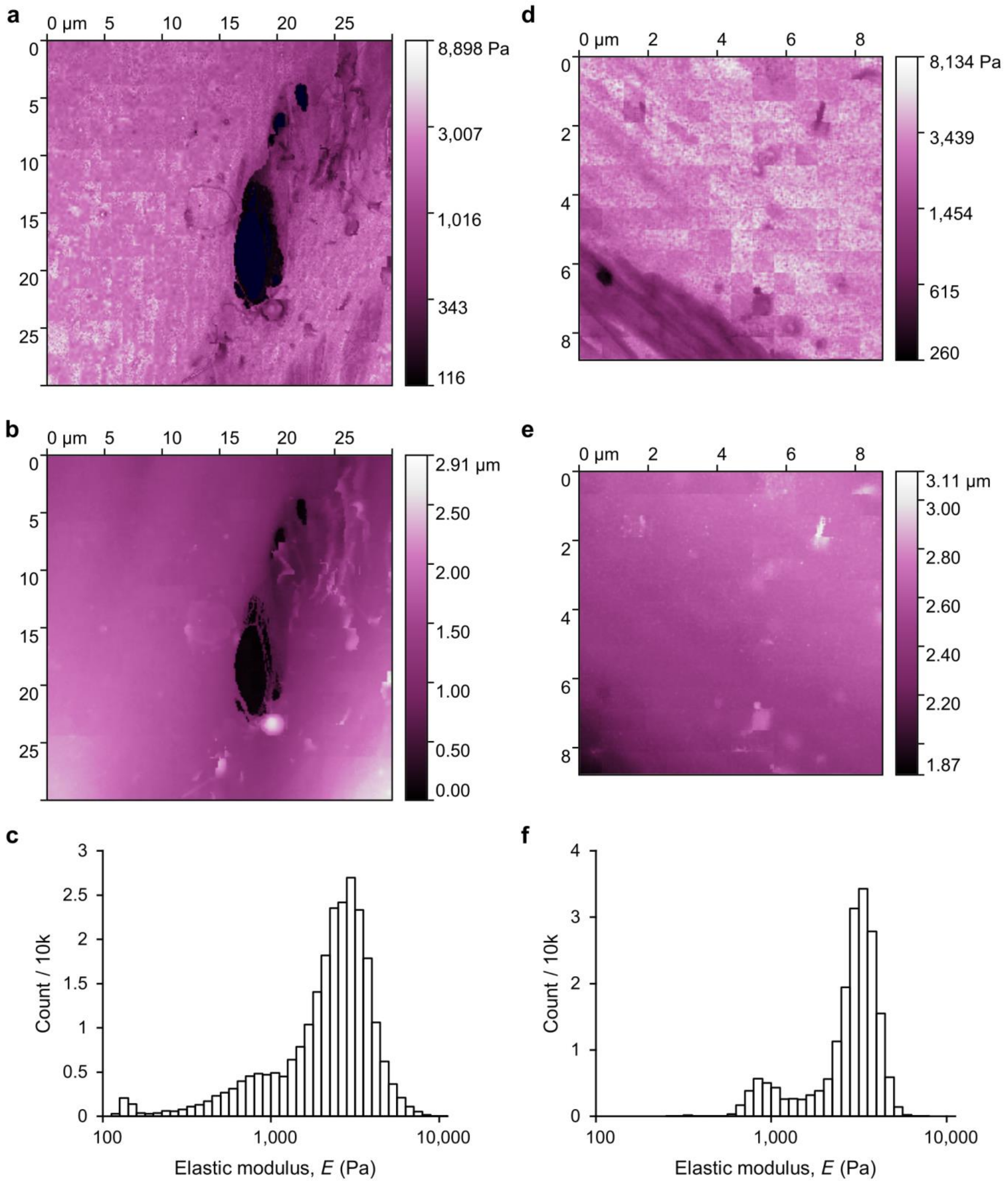

Figure S7. Maps of mammary fibroblast cell stiffness and topography. (**a**) Cell stiffness mapped with a 112 MΩ nanopipet of 80 nm aperture diameter at $\Delta I$=0.3%, 2.0%, exerting stress of 33.4 Pa, 151.9 Pa at gaps of 64.6 nm, 43.5 nm. (**b**) Topography. (**c**) Histogram of cell stiffness, selected as top 90% topography; mean 2.44 kPa, mode 2.82 kPa. (**d**) Higher resolution map of the stiffness of an 8.79 µm square area on a different cell mapped with a 90 MΩ nanopipet of 100 nm aperture diameter at $\Delta I$=0.3%, 3.0%, exerting stress of 0.4 Pa, 122.5 Pa at gaps of 81 nm, 49 nm. (**e**) Topography. (**f**) Histogram of stiffnesses, mean 2.89 kPa.

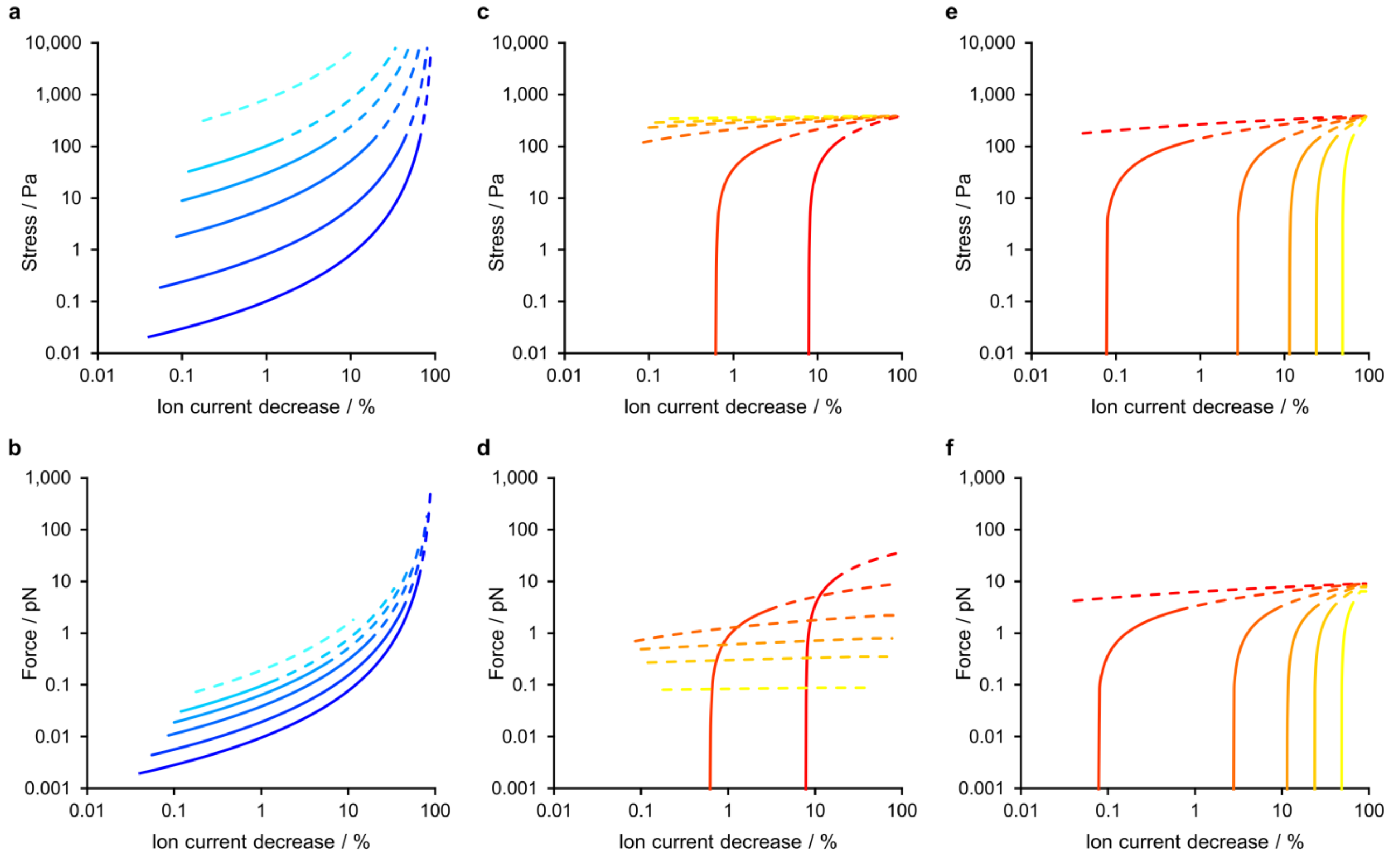

Figure S8. Stresses and forces versus ion current decrease calculated for ranges of aperture diameter and glycocalyx thickness. (**a**) Intrinsic stress for nanopipets of aperture diameters 10, 20, 30, 50, 100 and 200 nm, shown cyan to blue. The lines start at the minimum detectable ion current decrease in 1 ms and the dashes run from the bleb stress to the joining pressure. (**b**) Intrinsic force. Larger pipettes can apply a wider range of force before detaching any membrane because they detect the cell from further away and have larger tip-face area. (**c**) Direct stress on glycocalyx 70.5 nm thick for nanopipets of the same inner diameters, shown yellow to red. The smaller nanopipets would push the glycocalyx before a usable current decrease of 0.1%. (**d**) Direct force on this thickness of glycocalyx. (**e**) Direct stress exerted by a 100 nm aperture nanopipet for glycocalyx thicknesses of 10, 20, 30, 50, 100 and 200 nm, shown yellow to red. (**f**) Direct force on these thicknesses of glycocalyx; the asymptotes are equal here because the tip face areas are the same.